\documentclass[conference]{IEEEtran}
\usepackage{cite}
\usepackage{amsmath,amssymb,amsfonts}
\usepackage{algorithmic}
\usepackage{graphicx}
\usepackage{textcomp}
\usepackage{xcolor}
\usepackage{comment}
\usepackage{hyperref}
\usepackage{tabularx}
\usepackage{subcaption}
\usepackage[frozencache=true,cachedir=minted-cache]{minted}
\usepackage{flushend}
\usepackage{afterpage}

\hypersetup{hidelinks}

\def\BibTeX{{\rm B\kern-.05em{\sc i\kern-.025em b}\kern-.08em
    T\kern-.1667em\lower.7ex\hbox{E}\kern-.125emX}}

\newcommand{\rowtype}[1]{\ensuremath{\vec{#1}}}
\newcommand{\typecon}[1]{\ensuremath{\mathrm{#1}}}
\newcommand{\port}[1]{\ensuremath{\mathtt{#1}}}
\newcommand{\kw}[1]{\texttt{\small #1}}

\begin{document}

\title{Tierkreis: a Dataflow Framework for Hybrid Quantum-Classical Computing \\
}


\author{
  \IEEEauthorblockN{
    Seyon Sivarajah\IEEEauthorrefmark{1},
    Lukas Heidemann\IEEEauthorrefmark{1}\IEEEauthorrefmark{2}, 
    Alan Lawrence\IEEEauthorrefmark{1},  
    and
    Ross Duncan\IEEEauthorrefmark{1}\IEEEauthorrefmark{3}\IEEEauthorrefmark{4}
}
\IEEEauthorblockA{
  \IEEEauthorrefmark{1}Quantinuum, Terrington House, 13--15 Hills
  Road, Cambridge CB2 1NL, UK\\}

\IEEEauthorblockA{\IEEEauthorrefmark{2}Department of Computer Science,
  University of Oxford, 7 Parks Road, Oxford, OX1 3QG, UK\\}

\IEEEauthorblockA{\IEEEauthorrefmark{3}Department of Computer and
  Information Sciences, University   of Strathclyde, 26 Richmond
  Street, Glasgow, G1 1XH, UK\\}

\IEEEauthorblockA{\IEEEauthorrefmark{4}Department of Physics and
  Astronomy, University College London, Gower Street, London, WC1E
  6BT, UK\\} 
}

\maketitle

\begin{abstract}
  We present Tierkreis, a higher-order dataflow graph program
  representation and runtime designed for compositional,
  quantum-classical hybrid algorithms.  The design of the system is
  motivated by the remote nature of quantum computers, the need for
  hybrid algorithms to involve cloud and distributed computing, and
  the long-running nature of these algorithms.  The graph-based
  representation reflects how designers reason about and visualise
  algorithms, and allows automatic parallelism and asynchronicity.  A
  strong, static type system and higher-order semantics allow for high
  expressivity and compositionality in the program.  The flexible
  runtime protocol enables third-party developers to add functionality
  using any language or environment.  With Tierkreis, quantum software
  developers can easily build, visualise, verify, test, and
  debug complex hybrid workflows, and immediately deploy them to the
  cloud or a custom distributed environment.
\end{abstract}


\section{Introduction}
All quantum computing tasks involve the interplay between classical and quantum parts.
 The VQE algorithm  \cite{Peruzzo2014_temp} is typical: a classical driver program generates a set of quantum circuits; these circuits are submitted via a web API to a remote quantum computer; after waiting in a queue, the circuits are executed, and the results of these circuits are returned to the driver; when all the circuits complete, the results are the input to a classical optimiser, from which a new batch of circuits is generated, and the loop repeats.
 Another common pattern, seen in error-mitigation
 \cite{cirstoiu2022volumetric}, takes quantum circuits as inputs,
 modifies them, runs the modified circuits on a quantum device, and
 post-processes these results.  Fault-tolerant algorithms
 also often interleave quantum and classical steps
 \cite{Shor:PolyTimeFact:1997}.  The back-and-forth between classical
 and quantum phases, which run on different systems at different
 times, makes understanding, debugging, and verifying quantum software
 unnecessarily difficult.


 The classical phases of such algorithms may vary widely in their
 resource requirements.
 For example, a chemical simulation may require a
 memory-intensive integration calculation followed by storage of a
 large Hamiltonian, whilst GPU acceleration might be needed for a
 machine learning task.  Given the differing characteristics of 
 qubit technologies, some quantum devices will be better
 suited to particular algorithms -- or even subroutines -- than
 others.  From this perspective, quantum computers are special nodes
 in a larger heterogeneous distributed computing environment, and
 quantum software frameworks must allocate the right task to the right
 system.



 Quantum computers are intended to accelerate tasks which are
 classically infeasible, but that does not imply that the quantum
 algorithms are objectively fast.  On the contrary, when classical
 methods are impractical we expect that a quantum algorithm will be
 long-running.  Given that demand for quantum processors is likely to
 continue to exceed supply, we might also expect queuing and batch
 processing to remain the norm.  In this situation, tracking, pausing,
 and resuming the algorithm are valuable facilities, to monitor
 progress and recover from runtime failure.  Further, the high cost of machine
 time recommends the use of strong static analysis to detect
 programming errors before execution.  Few of these facilities are
 offered by current toolkits \cite{tket,Qiskit}.




 \emph{Tierkreis} is designed to address these needs. The
 program is represented as a dataflow graph where values flow along the edges,
 between nodes representing functional units, which potentially execute on different servers.
 Tierkreis has a strong static type
 system and supports higher-order functions which allow flexible
 control flow and compositional structures.

The system is designed to allow easy transition from existing quantum
computing workflows, which are predominantly made up of Python libraries and scripts,
to a paradigm which naturally mimics algorithm design, allows type-safety based
verification, and enables immediate deployment to cloud execution following
local testing and debugging.

Section \ref{sec:model} outlines the programming model of Tierkreis. The
runtime system is described in Section~\ref{sec:runtime}.

\subsection*{Related work}
Higher-order dataflow programming has been studied by many authors
\cite{bigdata,Fukunaga_Pree_Kimura_1993,towards}, albeit with
different semantics than Tierkreis, and the dataflow paradigm has been
applied to a variety of problem domains.  For example, the popular
machine-learning library TensorFlow \cite{tensorflow2015-whitepaper},
uses dataflow graphs of tensor values for optimised training of
machine-learning models;  LabVIEW \cite{labview} is a widely used
graphical environment for lab equipment testing and measurement; Enso
\cite{enso} is a functional language with both a visual and textual
representation, with a focus on data science visualisation and
analysis.  There are many others.  Our contribution in Tierkreis is
to define a semantics and execution model suitable for hybrid
quantum-classical computing, and which accommodates existing quantum
workflows, in a distributed cloud-native system.

There are other quantum software systems which aim to address similar
problems.  For instance, QCWare Forge and qBraid offer hosted python
environments alongside prebuilt libraries for executing quantum
experiments.  TensorFlow Quantum \cite{tf-quantum} allows adding
quantum steps to TensorFlow machine-learning workflows.  Qiskit
Runtime \cite{qiskit_runtime} enables offloading of some computation
to the IBM cloud, from where a hybrid computation involving IBM
quantum devices can execute.  Tierkreis and Qiskit Runtime are
complementary in design: Qiskit Runtime allows quantum subroutines to
be called as RESTful web services, whereas Tierkreis
can be used to build the whole program that would call such services.







\section{Programming Model}
\label{sec:model}
The program representation is designed with two primary goals: to naturally represent typical algorithms and to serve as a suitable runtime representation for a distributed, asynchronous execution model.
It is also intended to be a suitable compilation target for higher-level
languages or front-ends, such that they can be executed on the Tierkreis runtime.

\subsection{Dataflow}
\label{sec:dataflow}
The representation is a directed graph of \emph{nodes} connected by
\emph{edges}.  The nodes are labelled with \emph{function names} that identify pure functional
primitives: functions which compute a set of outputs from a set of
inputs, with no side-effects.  Edges connect nodes, from an
\emph{output port} of the source node to an \emph{input port} of the
target.  The port names identify the function inputs and outputs.  A
value produced at an output port flows along the edge to the target
input; values are immutable. Table~\ref{table:values} lists the available value types.
This structure is a \emph{dataflow graph} as it specifies only data dependencies:
a partial, rather than total, order of the operations. 
See Fig.~\ref{fig:zexp} for a visualisation of a simple Tierkreis graph, generated by the python package.  
This graphical approach is a natural conceptual model for a hybrid
algorithm -- with data moving between algorithmic units rather than as
a sequence of procedural steps.

\newcommand{\rulesep}{\unskip\ \vrule\ }
\begin{figure}[thb]
    \centering
    \begin{subfigure}[b]{0.45\linewidth}
      \centering
      \includegraphics[width=1\linewidth]{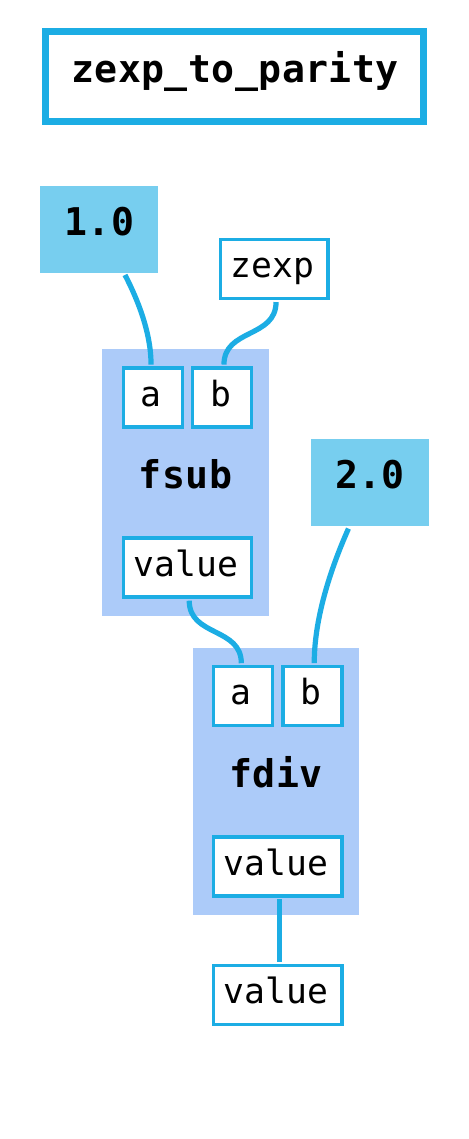}
      \caption{zexp\_to\_parity}
      \label{fig:zexp} 
    \end{subfigure}
    \hfill
    \rulesep
    \begin{subfigure}[b]{0.45\linewidth}
      \centering
      \includegraphics[width=1\linewidth]{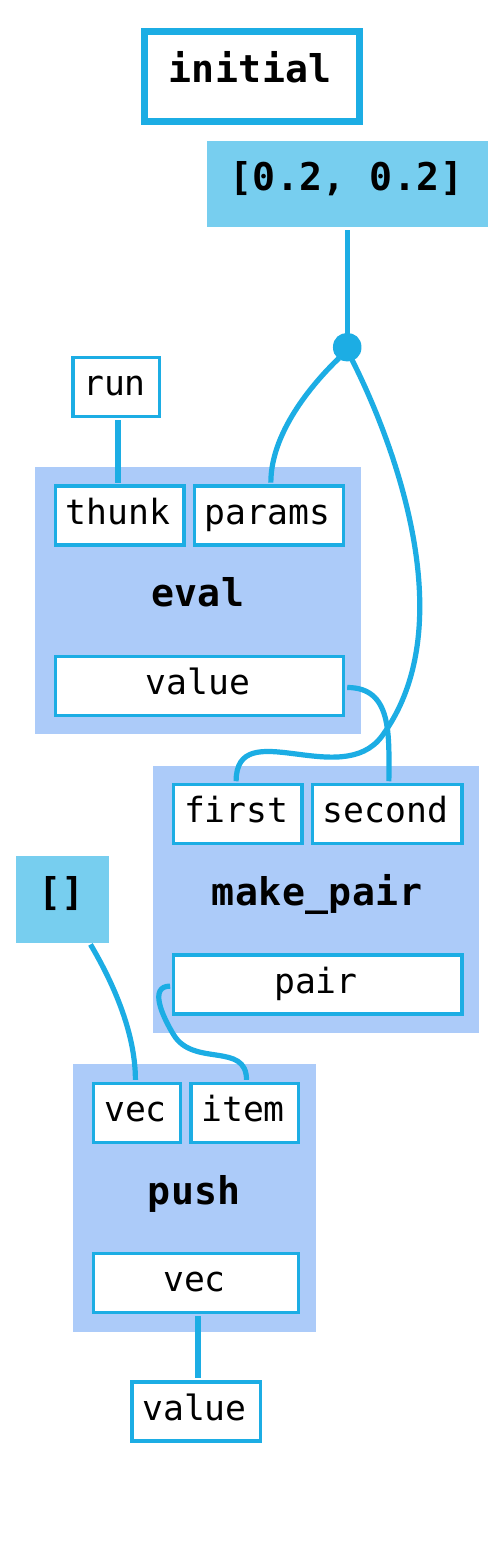}
      \caption{initial}
      \label{fig:initial} 
    \end{subfigure}
    \caption{Visualisations of simple Tierkreis graphs. Each box is a
      \emph{node} with wires denoting the \emph{edges} connecting
      them.  Function nodes are labelled with their names, and the names
      of their input (top) and output (bottom) ports; the inputs and
      outputs of the graph itself are shown in the same style.
      Constant nodes are labelled with their value.
      (a) A graph computing the function $f(x) = (1-x)/2$ on
      floating-point values.  The nodes labelled \kw{fsub} and
      \kw{fdiv} are built-in floating point subtraction and
      division functions respectively.
      (b) A graph which takes in a graph computing function $f$ via its
      \kw{run} port, and uses an \kw{eval} node to compute $f(x)$,
      where $x$ is the constant $[0.2,0.2]$ at the top.  We then use
      some built-in functions to construct $[(x,f(x))]$.  In
      Figure~\ref{fig:main} we show how this graph can be used used in
      a bigger program to compute the initial value for a variational
      optimising loop.
      }
    \label{fig:subfig}
\rule{\columnwidth}{0.5pt}
\end{figure}


\subsection{Asynchronous and parallel}

The approach also allows the program to be asynchronous and trivially
parallelisable without any further input from the programmer. Since nodes
correspond to pure functions, they can execute asynchronously
whenever their required inputs are ready.  Similarly any two nodes
whose inputs are available at the same time can execute in parallel,
at the discretion of the runtime, and could even execute on different
systems. The property that nodes can have multiple outputs (unlike
tree-like representations) allows the outputs of one node to
trigger multiple parallel nodes.

The error-mitigation routine Digital Zero Noise Extrapolation (ZNE)
\cite{zne} illustrates the use of parallel nodes.  Broadly, this error
mitigation scheme runs multiple versions of the input circuit on the
quantum device, artificially increasing the noise incurred each time.
In this implementation, using functions from the Qermit package
\cite{cirstoiu2022volumetric}, noise is increased by appending the
inverse of the circuit and then the circuit again, causing an odd
numbered multiplicative increase in circuit size (known as
\emph{folding}).  The resulting circuit is equivalent to the original,
but more noise is incurred due to additional operations.  Results from
each experiment, typically the expectation value of some operator, are
fitted to some function (e.g. linear, exponential) and the fit used to
extrapolate the ``zero-noise" result.

\begin{figure}
\centerline{\includegraphics[width=\linewidth]{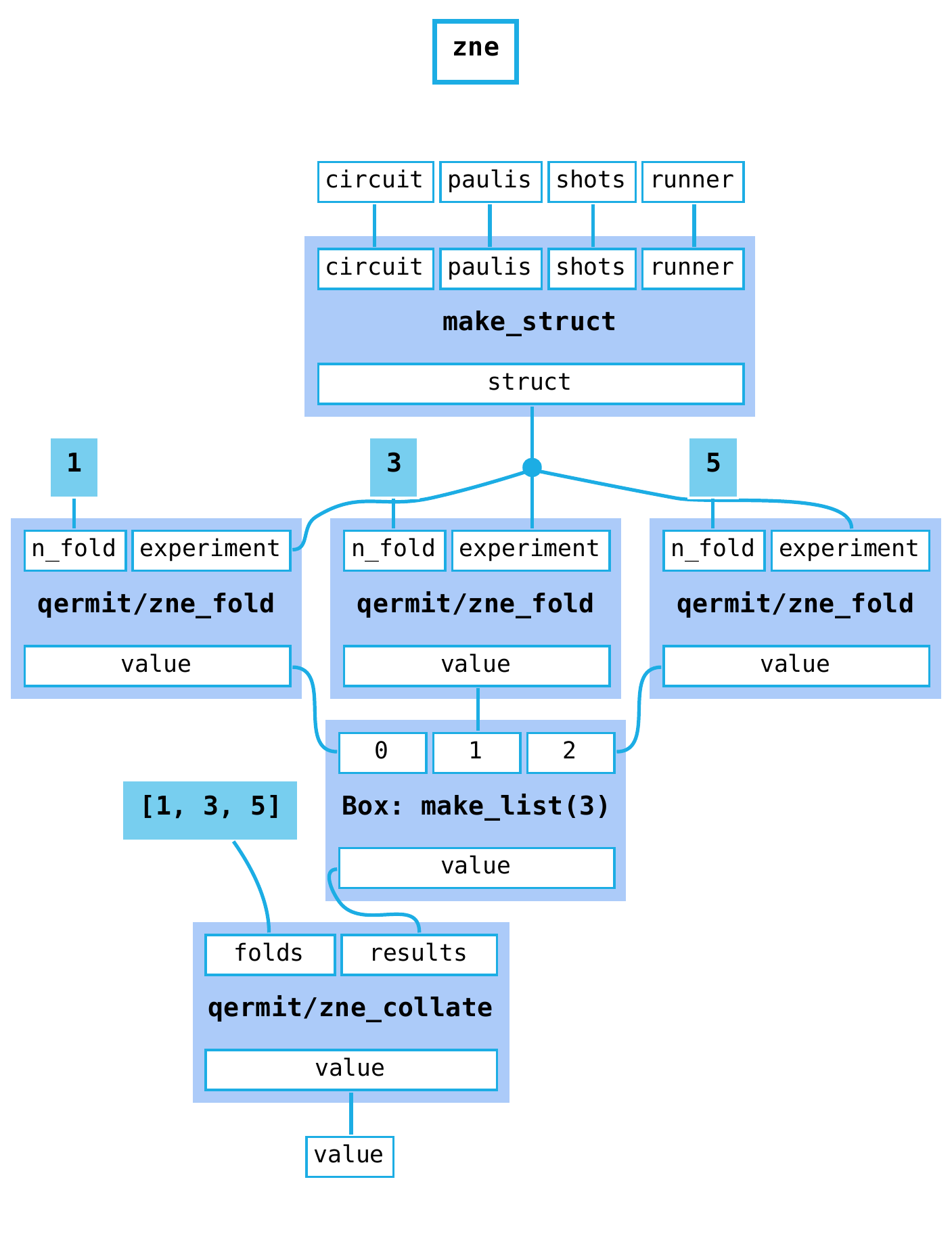}}
\caption{Tierkreis graph implementation of Zero Noise Extrapolation error mitigation, involving functions implemented using the Qermit package.
The spatially parallel placement of the \kw{zne\_fold} nodes naturally reflects the automatic execution parallelisation.}
\label{fig:zne}
\rule{\columnwidth}{0.5pt}
\end{figure}

The visualisation of the routine (Fig.~\ref{fig:zne}) shows that the
three \kw{zne\_fold} operations can execute in parallel, as they
each act on a copy of the data. The user need not make any additional
changes to their algorithm to parallelise it.

\subsection{Debug, Pause, Resume}
\label{sec:debug}

The dataflow approach also eases inspection and storage of the program \emph{state}. 
There is no program counter: instead, the unique value placed onto each
edge (if any) explicitly captures the entire state of the program.
This eases resuming from a checkpoint even after modifying the graph, e.g. when debugging a program.

\subsection{Higher-order}

\begin{figure}
    \centering
    \includegraphics[width=0.9\linewidth]{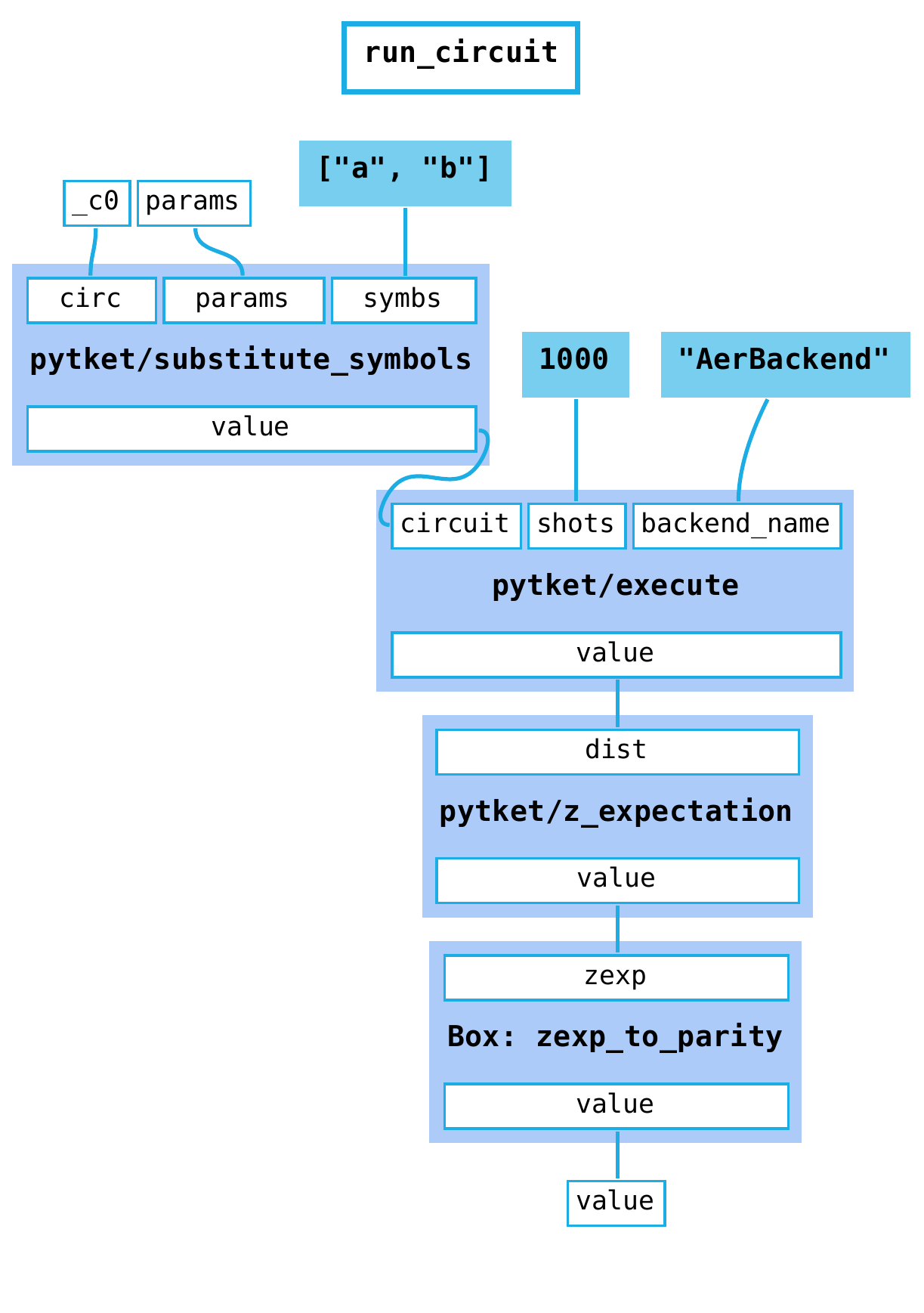}
    \caption{Tierkreis graph for calculating the cost function of a variational loop,
    using functions implemented using \kw{pytket}.
    The parameters in a symbolic circuit are set to those input, 
    the circuit is executed using a quantum device simulator,
    and the results used to calculate the expectation value of the parity of the measured distribution.
    The sub-graph contained in the \kw{zexp\_to\_parity} Box is shown in Fig.~\ref{fig:zexp}.
    The \kw{\_c0} input indicates the circuit has been automatically captured from 
    an outer scope. 
    At runtime a \emph{partial} node will insert the circuit as a constant prior to 
    the graph being executed, as shown in Fig.~\ref{fig:main}.
    }
    
    \label{fig:run_circuit}
\rule{\columnwidth}{0.5pt}
  \end{figure}

In Tierkreis, graphs are first-class values that can flow along edges,
and can be inputs or outputs of nodes.  This enables higher-order
operations such as evaluation, sequencing and partial application.
Graphs have the same interface as nodes -- named inputs and
outputs -- so the programmer can write complex algorithms in a modular
and \emph{compositional} style.  For example, most hybrid algorithms
wrap classical processing around an inner loop which executes a
quantum circuit.  Similarly, most error mitigation methods transform a
quantum execution in some way to achieve better results.  By providing
a graph for evaluating a circuit as a higher-order parameter, 
for example the \port{runner} input in
Fig.~\ref{fig:zne}, algorithms can use a basic circuit runner (like in
Fig.~\ref{fig:run_circuit}) or an error-mitigated alternative.  Indeed
the design of the Qermit \cite{cirstoiu2022volumetric} package is
based on composing mitigation schemes in just this manner.

In Tierkreis, values of \typecon{Graph} type are executed at runtime
using the built-in \kw{eval} function which, when provided with a graph
and suitable values for all its inputs, executes it, and returns the
output values.  Figure~\ref{fig:initial} shows an
example.  Another functional feature present in Tierkreis is partial
evaluation, via the \kw{partial} built-in.  Given a graph and some
of its inputs, \kw{partial} returns a new graph constructed by
plugging the inputs into the appropriate ports.

This feature is used by our Tierkreis front-end to automatically construct
\emph{closures}.  These allow input values which are in
scope at the definition point of the graph, to be retained at the
point of use. 
When building the program, the user can
use values from outer graph scopes freely, and at runtime these values
are inserted in to the inner scope using the \kw{partial}
function.  Figures~\ref{fig:run_circuit}~and~\ref{fig:loop_def} show
examples of graphs which capture outer scoped values, marked by the
\kw{\_c0} input port.  Fig.~\ref{fig:main} shows how these values
are injected by \kw{partial} nodes.

The graph also allows nesting subgraphs through the use of \emph{Box}
nodes. A Box node contains a subgraph and acts like a standard
function node with the same interface as the contained graph, akin to
a function invocation on a user-defined function.  This allows for
simple hierarchical structure and modularity in complex
algorithms. Fig.~\ref{fig:zne} contains the node user-labelled as ``\kw{Box:
  make\_list(3)}" signifying that it contains a subgraph rather than a
function.


\subsection{Control flow}

The execution model allows that a node may execute (only) when all of its inputs are ready;
this implies that a well-formed Tierkreis graph is
acyclic.  Despite this restriction, control flow constructs such as
iteration, branching, and pattern matching can be accommodated in
Tierkreis using special built-in higher-order functions and derived
constructs.



\begin{figure}
    \centering

    \includegraphics[width=1\linewidth]{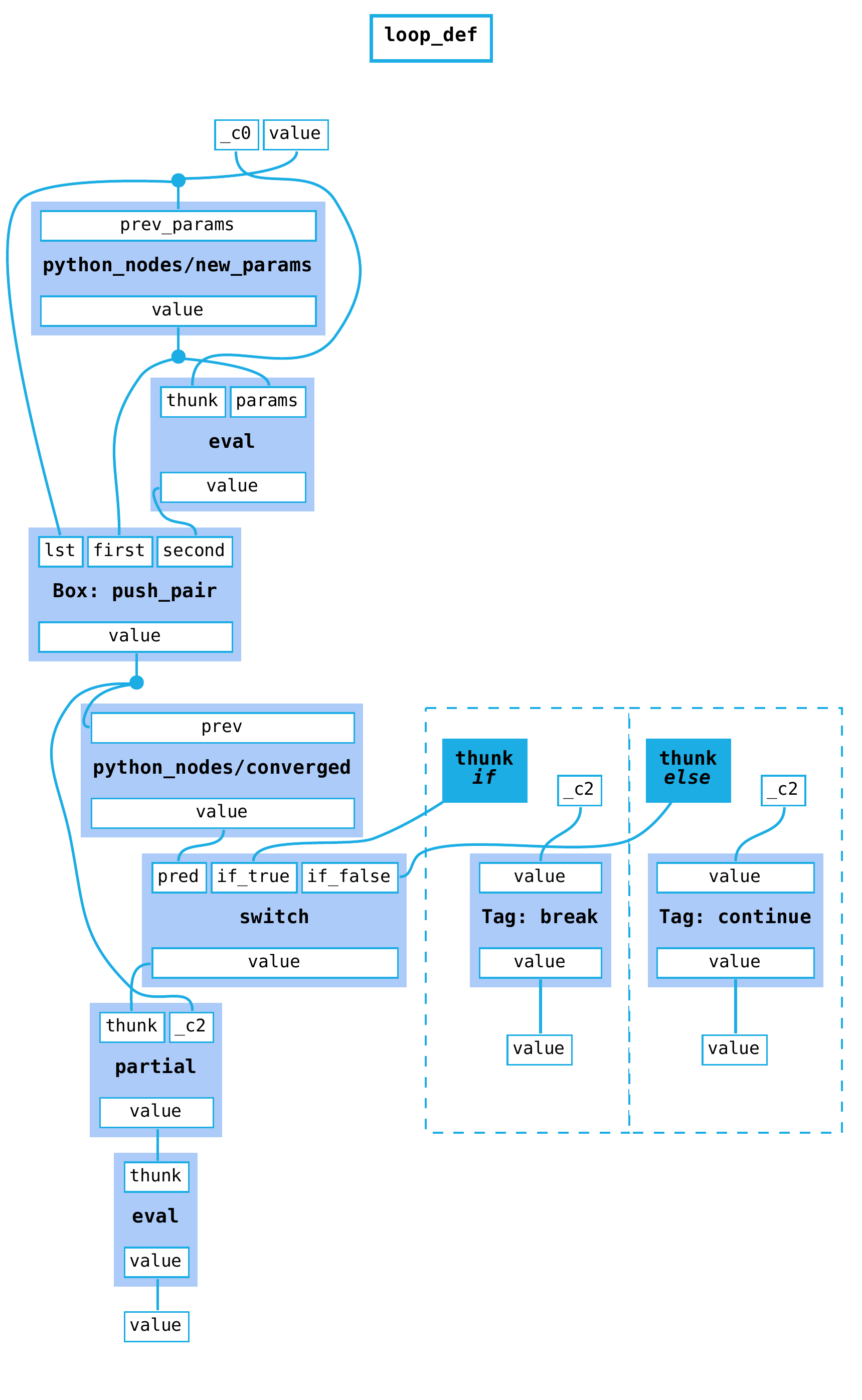}
    \caption{Tierkreis graph of a variational optimisation loop body.
    The worker implemented function \kw{new\_params} is used to 
    update parameter values, and \kw{converged} used to check
    if the optimisation is converged - signalling termination.
    The graph values \emph{if} and \emph{else} are shown 
    expanded inside dotted boxes.
        }
        \label{fig:loop_def}
\rule{\columnwidth}{0.5pt}        
\end{figure}

Branching is implemented via the \kw{switch} function, which takes
a boolean input and two others (with matching types), and returns one
of them based on the value of the boolean.  Using the \emph{switch}
with two graph values, and applying \kw{eval} to its return value
suffices to implement if-then-else.  (See Fig.~\ref{fig:loop_def} for
an example.)
Similarly, pattern matching on \emph{variants} is implemented using
the special \kw{Match} node, which takes a variant value as input
as well a graph to handle each variant of the type. \kw{Match}
chooses the appropriate graph value, and partially evaluates it with the inner
value held by the input variant.

Iteration is done via the \kw{loop} function, and a parameterised
\typecon{Variant} type with two tags, \kw{break} and \kw{continue}.
The \kw{loop} node accepts a \kw{body} graph and an initial
value, and returns an output.  The \kw{body} graph must accept the
initial value as its input, and return a value of the prescribed variant type.  If
it returns a value tagged \kw{continue}, this value is used as the
next input to \kw{body}; if it returns a value tagged
\kw{break}, this value is the output of the loop.  Our runtime implementations of
\kw{loop} use iteration, rather than recursively
unrolling the graph.

\begin{figure}
    \centering

    \includegraphics[width=0.8\linewidth]{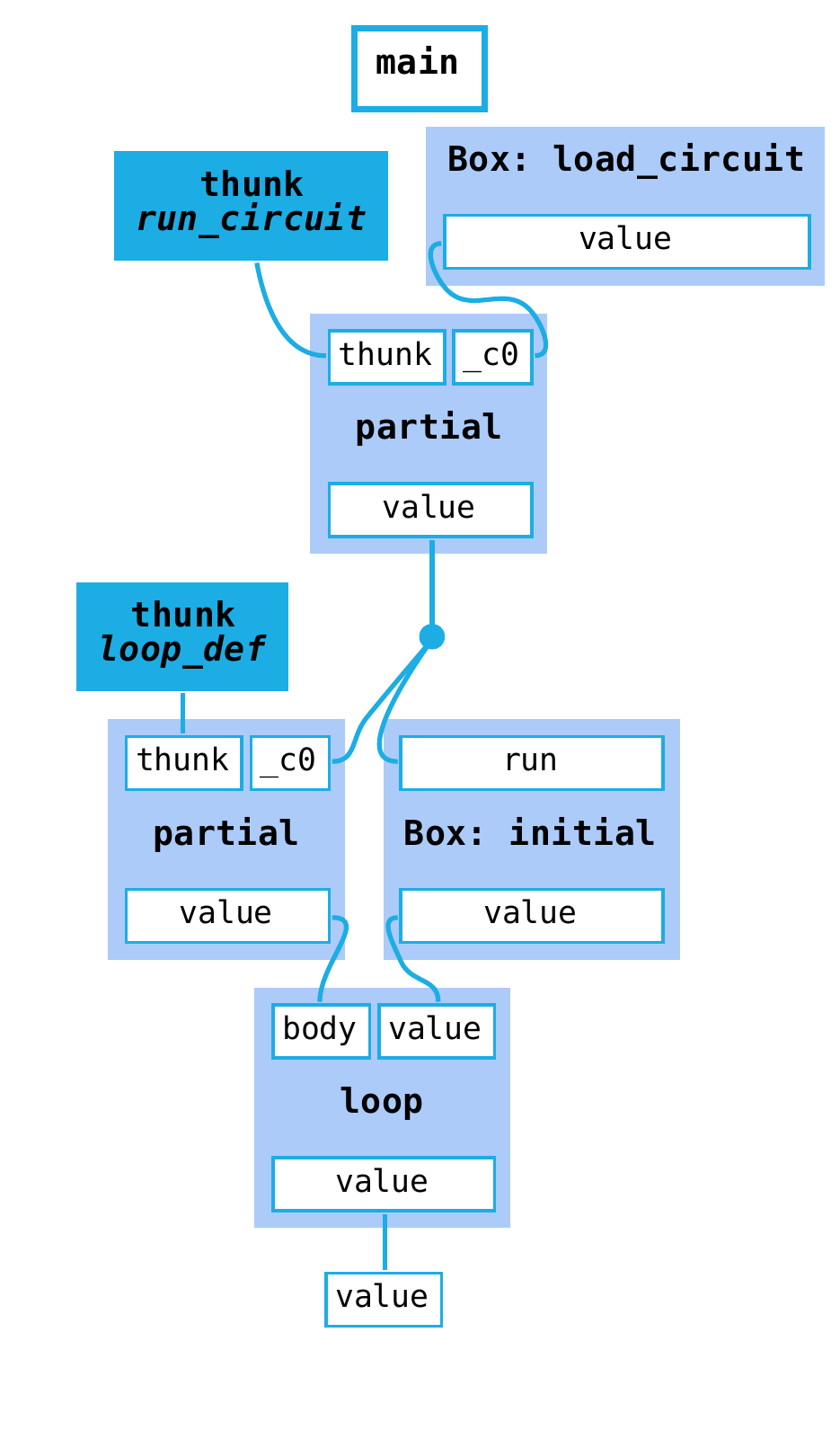}
    \caption{Tierkreis graph of a variational quantum algorithm.
    The graph values indicated by \kw{run\_circuit},
    \kw{loop\_def} and the box \kw{initial} can be seen in
    Figures~\ref{fig:run_circuit}, \ref{fig:loop_def} and \ref{fig:initial}
    respectively. Code for generating these graphs is listed in the Appendix.
        }
    \label{fig:main}
\rule{\columnwidth}{0.5pt}
\end{figure}

Fig.~\ref{fig:loop_def} gives an example of a loop body and 
Fig.~\ref{fig:main} shows how it is used with the \kw{loop} function.



\newlength\mylength
\setlength\mylength{\dimexpr.5\columnwidth-2\tabcolsep-0.5\arrayrulewidth\relax}

\begin{table}[t] 
\caption{Available value types in Tierkreis}
\renewcommand{\arraystretch}{1.3} 
\centering 
\begin{tabularx}{\columnwidth}{|l|X|}
\hline 
\bfseries Type & \bfseries Description\\ 
\hline\hline 
         
         \typecon{Bool} & True or False boolean.\\
        \hline 
     
         \typecon{Int} & 64-bit signed integer.\\
        \hline 
         
         \typecon{Float} & Double precision float.\\
        \hline 
         
         \typecon{Str} & Utf-8 encoded string.\\
        \hline 
         
         $\typecon{Pair}(A,B)$ & Pair (product type) of types $A, B$.\\
        \hline 
         
         $\typecon{Vec}(A)$ & Vector of element type $A$, with efficient \emph{push} and \emph{pop} operations.\\
        \hline 
         
         $\typecon{Map}(A, B)$ & Hash map of key type $A$ and value type $B$.\\
        \hline 
         
         $\typecon{Struct}(\rowtype{R})$ & Anonymous compound type with named fields and associated types.\\
        \hline 
         
  $\typecon{Variant}(\rowtype{R})$ & Tagged sum type.
                                     Also known as
                                     discriminated unions or
                                     enum types.\\ 
  \hline 
         
         $\typecon{Graph}(\rowtype{I} \rightarrow \rowtype{O})$ & A graph which takes inputs \rowtype{I} and returns outputs \rowtype{O}. \\

\hline 
\end{tabularx}
\label{table:values} 
\end{table}

\begin{table*}
\renewcommand{\arraystretch}{1.6} 
\caption{Some of the primitive operations available in Tierkreis.} 
\label{table_primitives} 
\centering 
\begin{tabularx}{\linewidth}{|l|p{9cm}|X|}
\hline 
\bfseries Operation & \bfseries Type & \bfseries Description \\ 
\hline\hline
    \texttt{eval} & $\forall \rowtype{I}, \rowtype{O}.\  (\port{thunk}: \typecon{Graph}(\rowtype{I} \rightarrow \rowtype{O}) \mid \rowtype{I}) \rightarrow \rowtype{O}$ & Evaluates a graph (determined at runtime) given values for all its inputs. \\
\hline
\texttt{partial} & $\forall \rowtype{I},\rowtype{I_1},\rowtype{I_2},\rowtype{O}.\ (\rowtype{I}_1 \sqcup \rowtype{I_2} \sim \rowtype{I}) \Rightarrow (\port{thunk}: \typecon{Graph}(\rowtype{I} \rightarrow \rowtype{O}) \mid \rowtype{I_1}) \rightarrow (\port{value}: \typecon{Graph}(\rowtype{I_2} \rightarrow \rowtype{O}))$ & Partially evaluates a graph given some inputs, returns a graph taking the remaining inputs. \\
\hline
\texttt{switch} & $(\port{pred}: \typecon{Bool}, \port{if\_true}:T, \port{if\_false}:T) \rightarrow (\port{value}:T)$ & Returns one of the \port{if\_true} / \port{if\_false} inputs according to \port{predicate}. \\
\hline
\texttt{push} & $\forall A. (\port{vec}: \typecon{Vec}(A), \port{item}: A) \rightarrow (\port{vec}: \typecon{Vec}(A))$ & Pushes an element onto a \typecon{Vector}. \\
\hline
\texttt{make\_pair} & $\forall A,B. (\port{first}: A, \port{second}: B) \rightarrow (\port{pair}: \typecon{Pair}(A,B))$ & Combines two elements into a \typecon{Pair}. \\
\hline
\texttt{fsub} & $(\port{a}: \typecon{Float}, \port{b}: \typecon{Float}) \rightarrow (\port{value}: \typecon{Float})$ & Floating point subtraction. \\
\hline
\texttt{fdiv} & $(\port{a}: \typecon{Float}, \port{b}: \typecon{Float}) \rightarrow (\port{value}: \typecon{Float})$ & Floating point division. \\
\hline
\texttt{make\_struct} & $\rowtype{R} \rightarrow (\port{struct}: \typecon{Struct}(\rowtype{R}))$ & Combines multiple values into a \typecon{Struct}. \\
\hline
\texttt{parallel} & $\forall \rowtype{I}, \rowtype{I}_1, \rowtype{I}_2, \rowtype{O}, \rowtype{O}_1, \rowtype{O}_2.\ (\rowtype{I}_1 \sqcup \rowtype{I}_2 \sim \rowtype{I}, \rowtype{O}_1 \sqcup \rowtype{O}_2 \sim \rowtype{O}) \Rightarrow (\port{a} : \typecon{Graph}(\rowtype{I}_1 \rightarrow \rowtype{O}_1), \port{b} : \typecon{Graph}(\rowtype{I}_2 \rightarrow \rowtype{O}_2)) \rightarrow (\port{value}: \typecon{Graph}(\rowtype{I} \rightarrow \rowtype{O}))$ & Combines two graphs with disjoint input and output ports into a single graph by composing in parallel. \\
\hline
\texttt{loop} & $\forall V,R. (\port{body}: \typecon{Graph}((\port{value}: V) \rightarrow (\port{value}: \typecon{Variant}(\port{continue}: V \mid \port{break}: R)), \port{value}: V)) \rightarrow (\port{value}: R)$ & Runs \port{body} until it returns \port{break}; possibly non-terminating. \\
\hline
\end{tabularx}
\\
\end{table*}

\subsection{Type system}
\label{sec:type}

\begin{figure}
    \centering

    \includegraphics[width=0.8\linewidth]{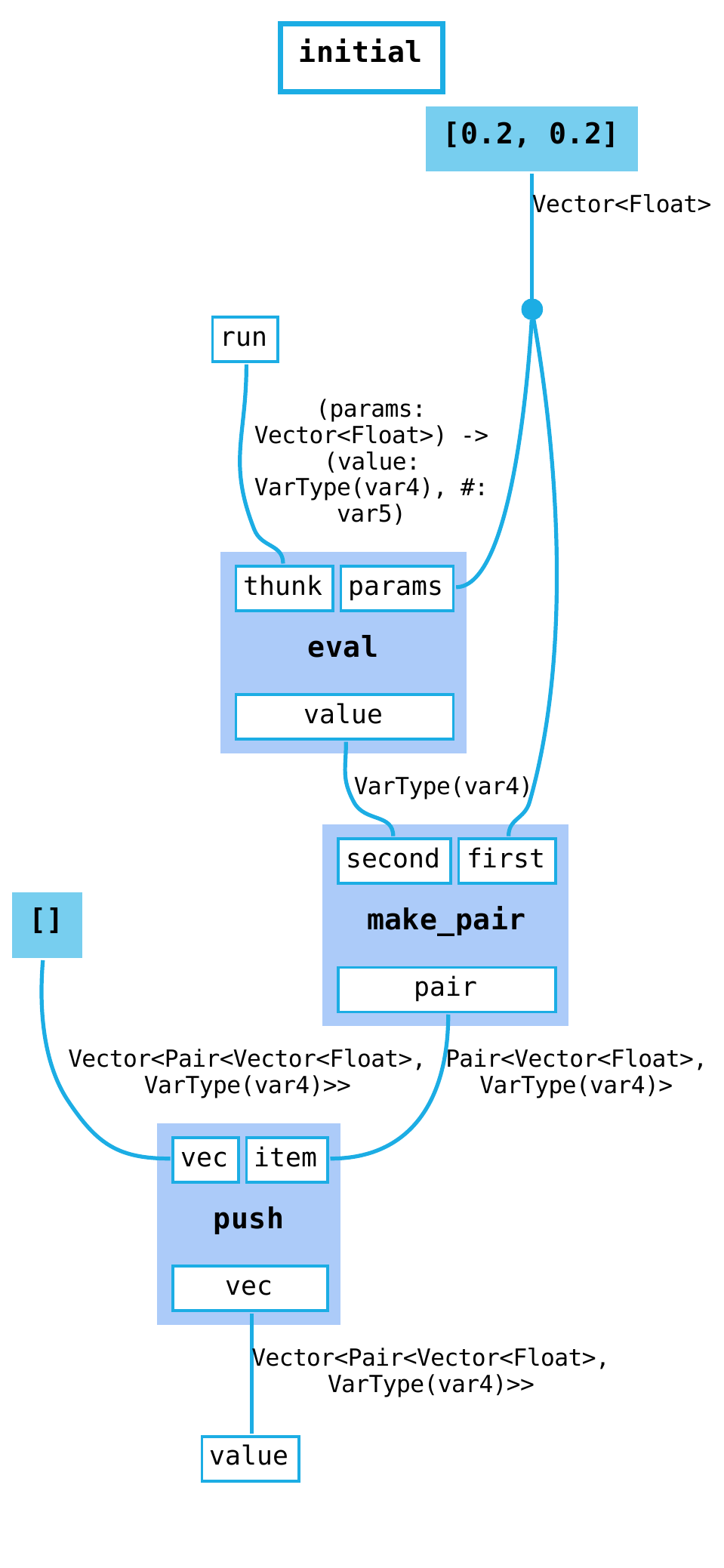}
    \caption{The graph from Fig.~\ref{fig:initial} with edge types annotated by the type system.
    The types of static values input to polymorphic functions constrain the inferred types of values
    on those edges. However, the return type of the input graph value \port{run} is not constrained,
    so the whole graph is polymorphic over the type variable \emph{var4}. 
    }
    \label{fig:initial_typed}
    \rule{\columnwidth}{0.5pt}
  \end{figure}

To improve the reliability of programs, and reduce the likelihood of runtime failure in the middle of a potentially costly or long-running computation, the Tierkreis graph has a strong static type system.
All values can be assigned a type from Table~\ref{table:values}, which includes user-definable algebraic data types via \typecon{Struct} and \typecon{Variant} types as well as \typecon{Graph} types for higher-order computation.
A graph is well-typed if, prior to runtime, type annotations  compatible with the signatures of all the nodes can be assigned to every edge.
Tierkreis implements a type inference algorithm which can fully infer the types of even completely unannotated programs, making user-provided annotations unnecessary. Fig.~\ref{fig:initial_typed} shows the graph from Fig.~\ref{fig:initial} with edge-annotations inferred by the algorithm.
In particular edges connected to the polymorphic functions \kw{eval, make\_pair}, and \kw{push} 
are annotated with concrete types when input constant values constrain them, and type variables when no such constraints exist, making the
whole graph polymorphic.

The signature of a function has a polymorphic type which is instantiated for every node that refers to the function.
The type inference algorithm then generates unification constraints between the types of input and output node ports which are connected by an edge.
After all constraints are collected Tierkreis attempts to find a most general type~\cite{garcia2015principal} which satisfies the constraints,
and returns the program with explicit type annotations filled in~\cite{pottier2014hindley}.
In the case of a type error Tierkreis provides high quality error messages by discovering a small set of locations which together make the program untypeable~\cite{loncaric2016practical}.

While the type system can statically guarantee that submitted programs compose operations in a type safe way, external workers
(see section~\ref{sec:workers})
could still produce outputs which do not conform to the type of the operation that they advertise.
To detect this early and provide the user with useful diagnostic information, the runtime system checks that the values produced by workers are consistent with the expected types.
The provided Python library for workers can additionally produce a Tierkreis type signature from Python type annotations to prevent a mismatch and to assist authors of worker functions in providing type safe operations.

A key component of Tierkreis' type system is the row type \rowtype{R}, which is a finite labelled list of types
denoted by $(\port{l}_1: \tau_1, \ldots, \port{l}_n: \tau_n)$. 
Rows are used to express the types of the input and output ports of nodes and \typecon{Graph} types, the fields of \typecon{Struct} types as well as the variants of \typecon{Variant} types.
Open rows $(\port{l}_1 : \tau_1, \ldots, \port{l}_n : \tau_n |\ \rowtype{R})$ allow for partial specification of a row type, where $\rowtype{R}$ is a type variable which stands in for the rest of the row.
This is used to give a type to operations such as \kw{eval} and \kw{make\_struct} (see Table~\ref{table_primitives}) which have a flexible set of input and output ports.
The labels of a row type are required to be distinct; to ensure that this property is respected for open rows, Tierkreis internally uses \emph{lacks constraints} $\port{l} / \rowtype{R}$ which prevent a row variable $\rowtype{R}$ from being unified with a row that contains the label $\port{l}$~\cite{gaster1996polymorphic}.
Lacks constraints are not shown in type signatures as they are implicit.
Partition constraints $\rowtype{R}_1 \sqcup \rowtype{R}_2 \sim \rowtype{R}_3$ can ensure that a row $\rowtype{R}_3$ is the disjoint union of rows $\rowtype{R}_1$ and $\rowtype{R}_2$, which is used to express the type of \kw{partial} application and \kw{parallel} composition of graphs.
In future work, rows will be used to tag computations with the resources they expect to be present in their environment, similar to the effect annotations used by Koka~\cite{leijen2013koka}.

\section{Runtime System}
\label{sec:runtime}

\begin{figure}
\begin{listing}[H]
\begin{minted}[fontsize=\footnotesize]{python}
nsp = Namespace("python_nodes")

@nsp.function()
async def exponent(a: float, b: float) -> float:
  return a ** b

if __name__ == "__main__":
  start_worker_server(Worker(), "test_worker", [nsp])
\end{minted}
\end{listing}
\caption{Worker implemented using the \kw{ tierkreis} Python package}
\label{fig:worker_impl}
\end{figure}

Whereas the Tierkreis graph defines the program to be run, the
Tierkreis runtime system is responsible for its execution.  The system
can be divided into two parts: the \emph{runtime}, which evaluates the
overall computation graph, and the \emph{workers} which provide
implementations of the individual nodes.

Once the client submits the graph and its required inputs, the runtime
need not interact with the client again until the output is ready,
making this a simple model for remote job submission.  While execution
must respect the data dependency specified in the graph structure,
task management is otherwise up to the implementation of the runtime.
In particular, it is free to execute the individual nodes concurrently
and asynchronously.

The runtime maintains an index of all the functions provided by its
workers, including the \emph{built-in} worker.  When a node is
triggered, the runtime invokes an implementation of matching signature
from its workers, passing the input values it received from the
predecessor nodes as arguments.  Should a node require a function not
provided by any of the workers, whether local or remote, the graph
cannot be executed. 
This results in a type-check error before any execution.

The Tierkreis system thus defines two interfaces to allow different
parts of the system to communicate: the \emph{runtime interface}
allows graphs to be submitted for execution, and the \emph{worker interface}
allows external function implementations to be called by
a runtime managing the execution.  Both interfaces are defined using
gRPC\footnote{https://grpc.io} and all data -- both types and values,
including graphs themselves -- are represented in Protocol Buffer
format\footnote{\url{https://developers.google.com/protocol-buffers/}}.
This provides many choices of implementation language and host
platform for Tierkreis clients and servers.

\subsection{Extensibility via Workers}
\label{sec:workers}
By implementing new workers, third-party developers can extend Tierkreis with arbitrary functions that can be used as node operations.  
Workers may be separate processes, communicating over gRPC, or in-process, serving function requests as sub-routines.
The worker interface can be asked to:
\begin{itemize}
\item list the functions it provides, as a \emph{signature} -— a mapping from names to type schemes, as in Table~\ref{table_primitives},
\item execute any of those functions given the inputs, returning the outputs.
\end{itemize}

Workers can be implemented in any language with a gRPC implementation;
the gRPC library does most of the work in setting up a server to
handle requests, leaving the programmer merely to implement the
functions.  
The \kw{tierkreis} Python package
automatically converts Tierkreis types
(Table~\ref{table:values}) to their Python equivalents, so a worker
can be as simple as in Figure~\ref{fig:worker_impl}.

Thus, workers allow the Tierkreis system to execute operations written in any language,
and provide a ``top down'' migration path: if only the outer high-level structure/control flow is ported to a Tierkreis graph,
it can call out to leaf functions written in legacy code (perhaps running on separate hardware).
A typical migration would involve implementing high-performance reusable algorithmic
blocks as worker functions using existing code, and porting the high-level algorithm by wiring these functions together in a graph.

Additionally a worker can call back into the runtime to execute a graph (and inputs)
and return the results back into the worker function; this allows a second ``bottom up'' migration route,
where top-level code in legacy languages can call leaf functions that have been ported to Tierkreis.


\subsection{Deployment}
\label{sec:deployment}

\begin{figure*}[thbp]
  \centering
  \[
    \begin{array}{lc}
      \text{(a)}  & \includegraphics[height=88pt]{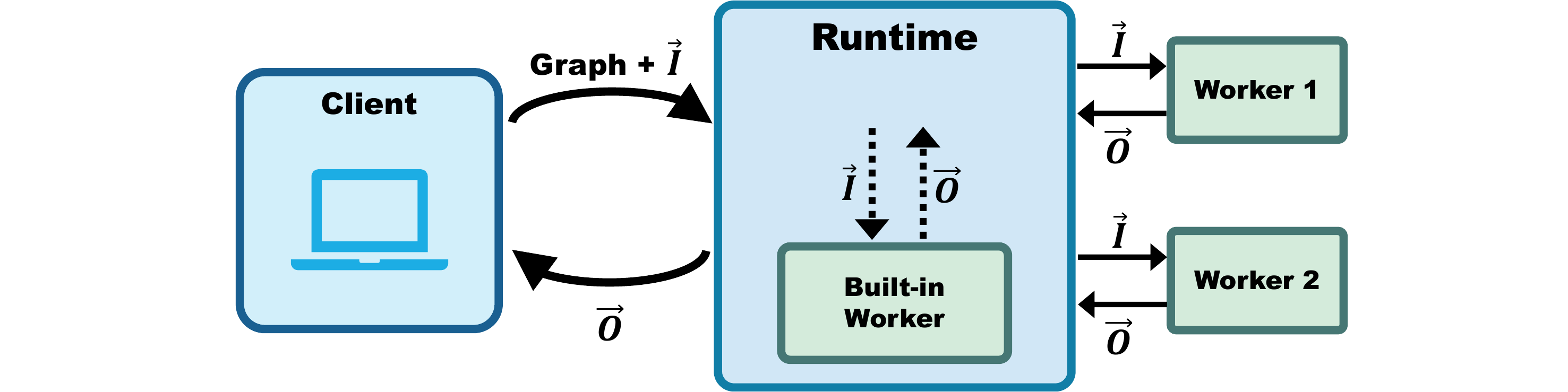} \\
      \\
      \text{(b)}  & \includegraphics[height=88pt]{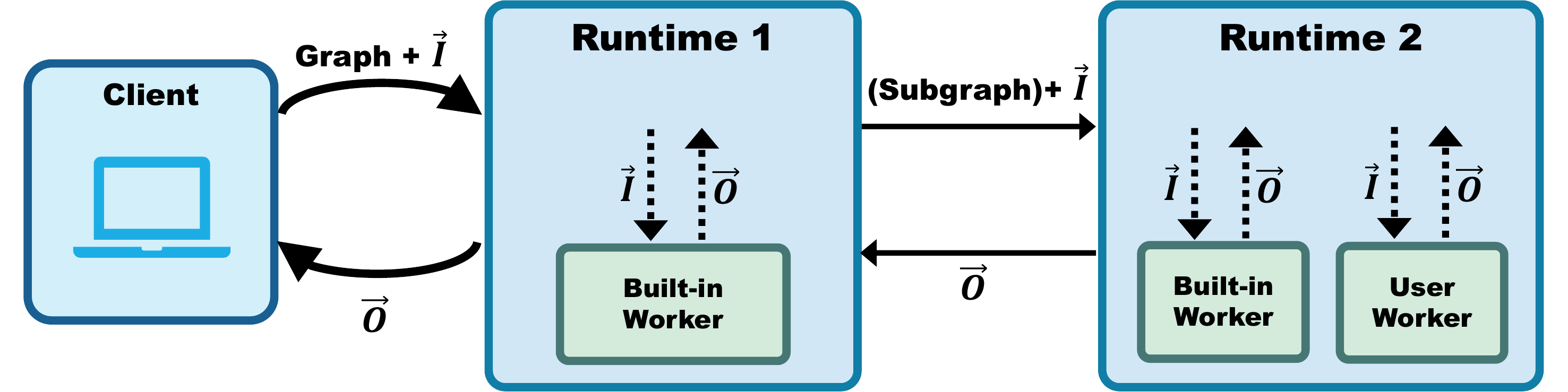} \\
    \end{array}
  \]
\caption{Two possible Tierkreis configurations.  Solid arrows indicate
  inter-process communication using protocol buffers and gRPC.  Dashed
  arrows indicate function calls within the same process.
  In (a) a local client
  interacts with a remote runtime, which uses two external workers to
  evaluate the nodes.  In (b) the runtime sends part of the graph
  to a second specialised runtime, which executes the entire subgraph
  and returns its output; this need not be visible to the client.  In
  both cases, the individual components can be located on different
  hosts; for example, co-located with the quantum hardware.}
\label{fig:runtimes}
\rule{\textwidth}{0.5pt}
\end{figure*}

A user uses the runtime interface to submit graphs,
typically over a network connection using some front-end client,
such as the one provided by our Python package.
In the simplest scenario, a client connects to a runtime,
which uses a local worker (in the same process) to provide a basic set of built-in functions
(some of which are shown in Table~\ref{table_primitives} above).
Since the state of each program is just the values placed onto edges so far,
this lends itself to a serverless implementation where every value set onto an edge is persisted into a store;
storing additional checkpoints (the start of each loop iteration, and optionally any currently-executing/pending requests to workers)
avoids redundant execution but is not essential.
However, a server where the state lives in memory may be preferred for performance reasons.
Fig.~\ref{fig:runtimes}(b) shows a schematic of such a setup.

At Quantinuum, we have two runtime implementations:
\begin{itemize}
    \item A Rust runtime, that runs as a gRPC service and supports connecting to arbitrary workers, also over gRPC.
    Since these workers are separate processes, they may run in independently sand-boxed environments,
    allowing them to be scaled and replicated as necessary, or run on dedicated hardware (e.g. for performance,
    functionality, security or IP protection reasons).
    This runtime is multi-threaded and fully asynchronous.
    \item A Python runtime, intended for testing and debugging and included in the Python package,
    that supports only workers written in Python (e.g. that of Fig.~\ref{fig:worker_impl}),
    and runs them directly in the same process as the runtime and the client code.
    This avoids the overhead of protobuf (de)serialisation to speed communication,
    which is particularly effective when large volumes are data are transmitted between worker functions.
\end{itemize}

The Python runtime allows inspection of intermediate values in the graph via a user-provided callback function.
A future extension to the Rust runtime will allow intermediate values to be streamed back to the client as they
appear on edges.

\subsection{Tree of runtimes}

The runtime interface includes the worker interface, with its
signature being the union of the signatures of the
workers. Hence, a runtime acting as a client can connect to
another acting as a server, just like any other worker. This enables
a tree of distributed runtimes, with parents requesting children to execute functions; a runtime receiving such a request forwards it onto an appropriate one of its children. 
Ongoing work will allow a parent runtime to send a subgraph (marked by a \emph{Box}) for execution on a child runtime, allowing data local to that subgraph to stay on the child runtime for performance.
For example, there could be a runtime instance co-located to the quantum device, that runs the variational loop subroutine of a larger algorithm to reduce latency, while the parent runtime in the cloud (with which the client interacts) runs the rest. A schematic of such a setup is in Fig.~\ref{fig:runtimes}(b).




\section{Conclusion}
\label{sec:conclusion}
Tierkreis is a powerful system for programming cloud-native
quantum-classical algorithms.  The dataflow graph allows higher-order
algorithms to be represented naturally, and, when combined with the
runtime, unlocks automatic parallelism, asynchronous execution and
distribution over the cloud. Platform-neutrality (via protobuf)
and extensibility via workers allows easy and graduated migration of existing code as well
as composition of components from different languages and environments.

The \kw{tierkreis} Python package along with all gRPC and Protocol
Buffer definitions are available under an open-source 
licence\footnote{\url{https://github.com/CQCL/tierkreis}}.


\section*{Acknowledgments}

The authors would like to thank Conor McBride and Craig Roy for
invaluable discussions on the design of Tierkreis, and John Children
for also kick-starting much of the implementation.  We also thank Alex
Chernoguzov, William Whistler and Alex Rice for their helpful feedback
on earlier drafts.

\enlargethispage{-0.7cm} 
\bibliography{IEEEabrv, cites}

\clearpage
\onecolumn


\begin{listing}[!htp]
\begin{minted}{python}
a, b = symbols("a b")
ansatz = Circuit(2)
ansatz.Rx(0.5 + a, 0).Rx(-0.5 + b, 1).CZ(0, 1).Rx(0.5 + b, 0).Rx(
    -0.5 + a, 1
).measure_all()

@graph()
def push_pair(lst, first, second) -> Output:
    pair = bi.make_pair(first, second)
    return Output(bi.push(lst, pair))

push_pair_box = Box(push_pair())

@graph(sig=sig)
def initial(run) -> Output:
    init_params = Copyable(Const([0.2, 0.2]))
    init_score = bi.eval(run, params=init_params)
    return Output(push_pair_box(Const([]), init_params, init_score))

init_box = Box(initial())

@graph()
def load_circuit() -> Output:
    js_str = json.dumps(ansatz.to_dict())
    c = pt.load_circuit_json(Const(js_str))
    return Output(c)

@graph()
def zexp_to_parity(zexp) -> Output:
    y = bi.fsub(Const(1.0), zexp)
    return Output(bi.fdiv(y, Const(2.0)))

@graph()
def main() -> Output:
    circ = Box(load_circuit())()

    @closure()
    def run_circuit(params: Input) -> Output:
        syms = Const(["a", "b"])
        # substitute parameters in circuit with values a, b
        subs = pt.substitute_symbols(circ, syms, params)
        res = pt.execute(subs, Const(1000), Const("AerBackend"))
        o = Output(Box(zexp_to_parity())(pt.z_expectation(res)))
        return o

    run_circuit.copyable()

    @loop()
    def loop_def(initial: Input[Any]) -> Output:
        recs = Copyable(initial)
        new_cand = Copyable(pn.new_params(recs))
        score = run_circuit(new_cand)
        recs = Copyable(push_pair_box(recs, new_cand, score))

        with IfElse(pn.converged(recs)) as lbody:
            with If():
                Break(recs)
            with Else():
                Continue(recs)
        o = Output(lbody.nref)
        return o

    init_val = init_box(run_circuit.graph_src)
    return Output(loop_def(init_val))
\end{minted}
\caption{{Python 3 source for building a graph for a simple variational algorithm. 
A symbolic \kw{pytket} \cite{tket} quantum circuit is built and inserted as a static constant.
Features demonstrated include decorator-based builder environments, automatic capture of values in closures,  and incremental type inference (if the addition of a node fails an incremental type check, a Python exception is raised at the offending line).
Functions from the \emph{builtin} (\kw{bi}), \emph{python\_nodes} (\kw{pn}) and \emph{pytket} (\kw{pt}) workers are used.
Figures~\ref{fig:subfig}, \ref{fig:zne}, \ref{fig:run_circuit}, \ref{fig:loop_def}, and \ref{fig:main} show visualisations of the 
graphs defined here. Imports and other setup code are not included for brevity.
}}
\label{lst:build}
\end{listing}

\clearpage
\onecolumn
\begin{listing}[!htp]
\begin{minted}{python}
@namespace.function()
async def new_params(
    prev_params: list[CandRecord],
) -> list[float]:
    assert len(prev_params) > 0
    if len(prev_params) == 1:
        return [random.random()] * len(prev_params[0])

    first, second = prev_params[-2:]

    ar1 = np.array(first[0])
    ar2 = np.array(second[0])

    diff = ar2 - ar1
    diff[diff == 0.0] = 1e-14
    grad = (second[1] - first[1]) / diff
    ar2 -= 0.1 * grad
    return list(ar2)


@namespace.function()
async def converged(prev: list[CandRecord]) -> bool:
    n = 5
    if len(prev) >= n:
        last_n = np.array([x[1] for x in prev[-n:]])
        moving_av = sum((last_n - last_n.mean()) ** 2)
        return bool(moving_av < 1e-5)

    return False
\end{minted}
\caption{
Worker definitions of gradient descent functions used in the example,
using the worker building tools from the Tierkreis Python package.
}

\label{lst:opt_worker}
\end{listing}

\end{document}